# Ultrafast internal dynamics of chiral domain walls probed by time-resolved XRMS

Dmitriy Ksenzov[1], Flavio Capotondi[2], Nicolas Jaouen[3,4], Raphael Gruber[5], Vincent Cros[6], Mathias Kläui[5], Matteo Pancaldi[2,3], Emanuele Pedersoli[2], Nicolas Reyren[6], and Christian Gutt[1].

[1] *Universität Siegen, Walter-Flex-Straße 3, 57072 Siegen, Germany*

[2] *Elettra Sincrotrone Trieste, Strada Statale 14, km 163.5, 34149 Basovizza, TS, Italy*

[3] *Department of Molecular Sciences and Nanosystems, Ca' Foscari University of Venice, Venice, Italy*

[4] *Synchrotron SOLEIL, Saint-Aubin, Boîte Postale 48, 91192 Gif-sur-Yvette Cedex, France*

[5] *Institut für Physik, Johannes Gutenberg-Universität Mainz, 55099 Mainz, Germany*

[6] *Laboratoire Albert Fert, CNRS, Thales, Université Paris-Saclay, 91767 Palaiseau, France*

We measure picosecond dynamics of labyrinthine stripe domains with chiral Néel-type domain walls using time-resolved x-ray resonant magnetic scattering (XRMS). At the stripe-domain wavevector, the helicity-summed signal shows ultrafast demagnetization, recovery, and a weak oscillatory contribution that we identify as a signature of laser-launched coherent surface phonons. In contrast, the dichroic signal, which is sensitive to the in-plane magnetization inside the Néel walls, shows a strong oscillatory response whose frequency decreases with increasing pump fluence. We attribute this softening to pump-induced changes in the effective anisotropy and saturation magnetization, which modify the restoring field of an internal domain-wall mode. Time-resolved XRMS thus isolates wall-specific dynamics and provides access to internal domain-wall motion in disordered stripe textures on picosecond time scales.

Magnetic multilayers with perpendicular magnetic anisotropy (PMA) can host chiral Néel domain walls stabilized by the interfacial Dzyaloshinskii-Moriya interaction (i-DMI) and support current-driven domain-wall motion [1–3]. The same interactions can also stabilize topological spin textures such as skyrmions in multilayer systems [4,5].

Ultrafast optical excitation can modify magnetic order on sub-picosecond time scales, but the picosecond dynamics of chiral domain walls remain poorly explored experimentally. This regime is particularly relevant when transient heating drives the system toward the Curie temperature, softens the effective anisotropy, and modifies the domain-wall structure and stiffness [6,7].

Femtosecond optical pump pulses generate transient strain, change the magnetic free energy, and perturb the internal wall structure. Time-resolved magneto-optical probes often mix these responses and do not isolate the reciprocal-space signature of a domain pattern. In contrast, time-resolved x-ray resonant magnetic scattering (XRMS) probes

Fourier components of magnetic textures and separates non-dichroic and dichroic signals through helicity summation and helicity difference [8–10]. Ultrafast measurements that isolate chiral domain-wall dynamics remain limited [11,12].

In this study, we use time-resolved XRMS to measure the picosecond response of chiral domain walls after femtosecond optical pumping. By tracking scattering at the domain wavevector, we distinguish a weak structural oscillation from the magnetic response and identify a fluence-dependent oscillation in the dichroic signal, which we discuss below in terms of internal domain-wall dynamics [1,12,13].

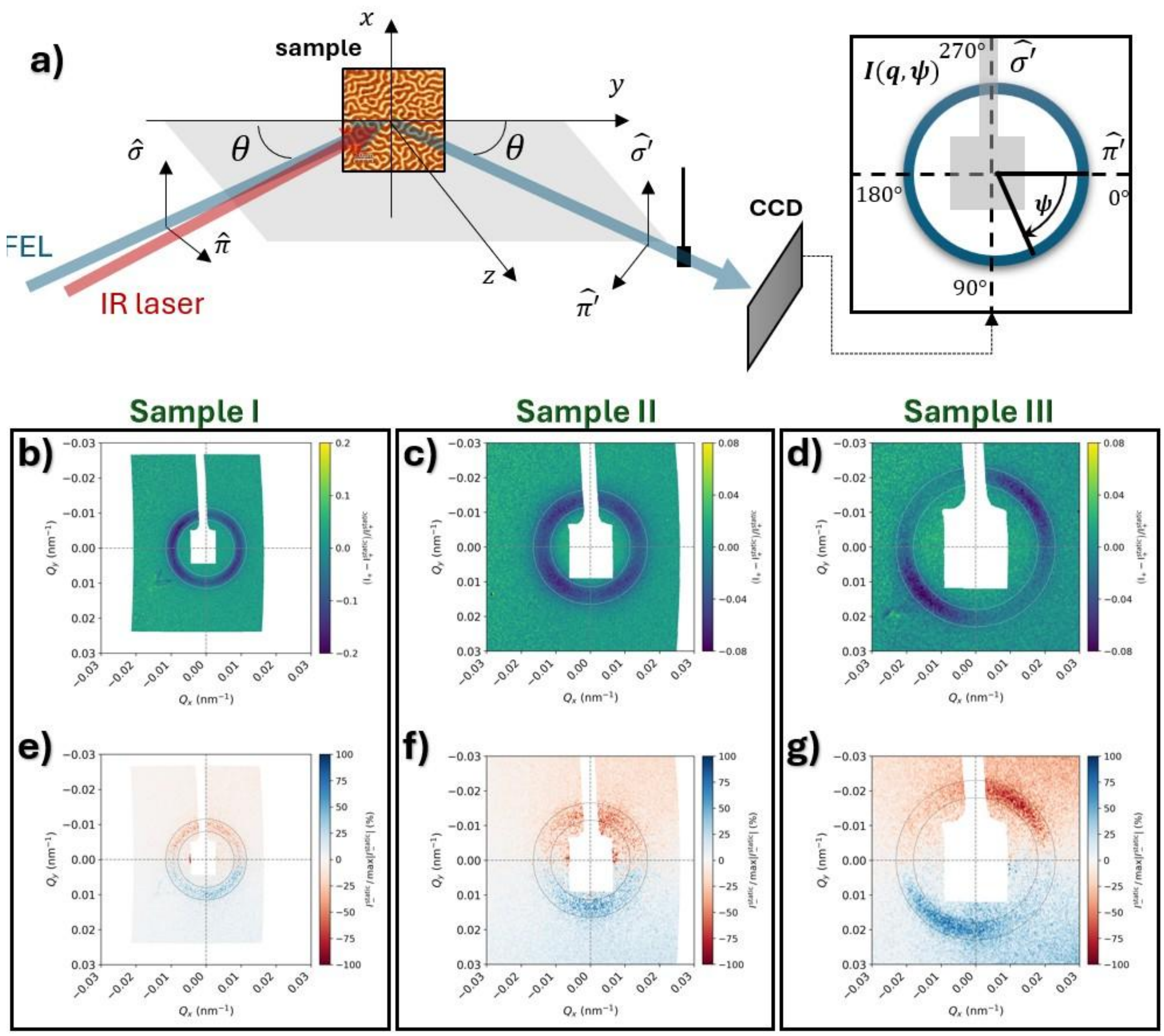

FIG 1: **(a)** Schematic of the time-resolved XRMS experiment. Infrared (IR) pump pulses ($\lambda_L = 790\ nm$) excite the multilayer. Circularly polarized FEL pulses tuned to the Fe $M_{2,3}$ edge ($\lambda = 22.75\ nm$) probe the response in reflection geometry at $\theta \approx 45°$. The sample-to-detector distance was 150 mm for sample I, 100 mm for sample II, and 75 mm for sample III. **(b-d)** Normalized differential sum signal, $\left(I_+^{pumped} - I_+^{static}\right)/I_+^{static}$, at a pump-probe delay of 0.5 ps for three Si/$SiO_2$/$[HM/CoFeB/MgO]_n$ /Ta multilayers: sample I, Ir(6.6 nm)/CoFeB(0.7 nm)/MgO(2.0 nm) x15; sample II, W(5.0 nm)/CoFeB(0.8 nm)/MgO(2.0 nm) x7; and sample III, Ir(6.6 nm)/CoFeB(0.7 nm)/MgO(2.0 nm) x7. The scattering ring corresponds to the mean stripe-domain periodicity; its intensity decreases after pumping due to reduced magnetic scattering contrast. (**e-g**) Normalized static dichroic signal, $I_-^{static}$, sensitive to in-plane domain-wall magnetization. According to the scattering geometry [panel (a)], samples I and II show Néel-type walls, while sample III exhibits mixed Néel and Bloch character.

Time-resolved soft XRMS experiments were performed at the DiProI endstation of the FERMI free-electron laser (FEL) facility in Trieste, Italy [14-16]. A schematic of the experimental geometry is shown in Fig. 1(a). The magnetic multilayers were measured in reflection geometry at an incidence angle of $\theta = 45°$. Ultrafast excitation was provided by a nearly collinear infrared pump pulse ($\lambda_L$ = 790 nm) with linear horizontal (LH) polarization and a pulse duration of ~60 fs. The absorbed pump fluence *F* varied between 1.5 and 4.5 mJ $cm^{-2}$. The pump spot on the sample surface was ~600 x 670 $\mu m^2$.

The FEL probe pulses had a duration of approximately 60 fs and were delivered at a repetition rate of 50 Hz. The probe wavelength was 22.75 nm, corresponding to resonant sensitivity at the Fe $M_{2,3}$ edge. The probe beam was focused to a spot size of approximately 320 x 290 $\mu m^2$. The pump-probe delay *t* was varied up to 1 ns using an optical delay stage. Scattered extreme-ultraviolet radiation was recorded in reflection geometry using a two-dimensional charge-coupled-device (CCD) detector, placed at 150, 100, and 75 mm from samples I, II, and III, respectively; the specular beam was blocked by a beamstop. At each delay, patterns were averaged over 500 pump–probe shots.

Measurements were performed on three Si/SiO2/$[HM/CoFeB/MgO]_n$/Ta multilayers, where HM denotes the heavy-metal underlayer and the individual layer thicknesses are specified in the caption of Fig. 1. Sample I was an Ir-based multilayer with n = 15, sample II was a W-based multilayer with n = 7, and sample III was an Ir-based multilayer with n = 7. All multilayers exhibited PMA and formed labyrinthine stripe domains in the remanent state at room temperature (~300 K). The stripe-domain periodicity was determined from the radius of the magnetic scattering ring. The domain wave vectors were $Q_M \approx 0.010$ $nm^{-1}$ (sample I), $Q_M \approx 0.0145$ $nm^{-1}$ (sample II), and $Q_M \approx 0.021$ $nm^{-1}$ (sample III), corresponding to stripe-domain periods of ~630, 430, and 300 nm, respectively.

In resonant magnetic scattering, the total scattering amplitude is written as the sum of structural and magnetic terms [8],

$$f = f_0 + f_m, \tag{1}$$

where $f_0$ and $f_m$ are the charge and magnetic scattering amplitudes. Using left- and right-circularly polarized probe pulses, the measured intensities are $I_{CL}$ and $I_{CR}$. We define the helicity-summed and dichroic signals as

$$I_+ = I_{CL} + I_{CR}, \quad I_- = I_{CL} - I_{CR} \tag{2}$$

To leading order [8, 9],

$$I_+ \propto |f_0|^2 + |f_m|^2, \quad I_- \propto \Im m(f_0 f_m^*), \tag{3}$$

where the dichroic term arises from charge-magnetic interference. In the present reflection geometry with stripe domains, $I_-$ is primarily sensitive to the in-plane magnetization associated with Néel-type domain walls. Bloch-type domain walls also contribute to the dichroic signal, but with a different azimuthal symmetry, allowing their contribution to be distinguished and treated as a small correction, which is neglected here [10].

Figs. 1(b)-1(d) show the normalized differential sum signal $\Delta I_+/I_+^{static}$, defined as $\left(I_+^{pumped}(t) - I_+^{static}\right)/I_+^{static}$, at *t* = 0.5 ps for samples I-III. The reduced ring intensity at $Q_M$ is consistent with reduced magnetic scattering intensity due to optically induced

ultrafast demagnetization [7,11,17]. The dichroic signal is sensitive to the chiral magnetic contribution. The static dichroic patterns in Figs. 1(e)-1(g), $I_-^{static}/\max|I_-^{static}|$, show an angular modulation set by the in-plane domain-wall magnetization and its Fourier components at $Q_M$ [9,18].

For samples I and II, the maxima of dichroic contrast in the static, unpumped remanent state occur at 90° and 270°, consistent with Néel-type walls with a unique chirality throughout the multilayers. In sample III, the extrema are shifted to 135° and 315°, indicating a finite Bloch admixture and thus mixed Néel–Bloch character of the domain walls [5,9,19].

Fig. 2 summarizes $\Delta I_+/I_+^{static}$ at the magnetic ring associated with $Q_M$. The traces were obtained by averaging over $Q_M$ in [0.0195, 0.0215] nm$^{-1}$ for sample III (Figs. 2(a) and 2(c)) and over $Q_M$ in [0.0092, 0.0101] nm$^{-1}$ for sample I (Figs 2(b) and 2(d)).

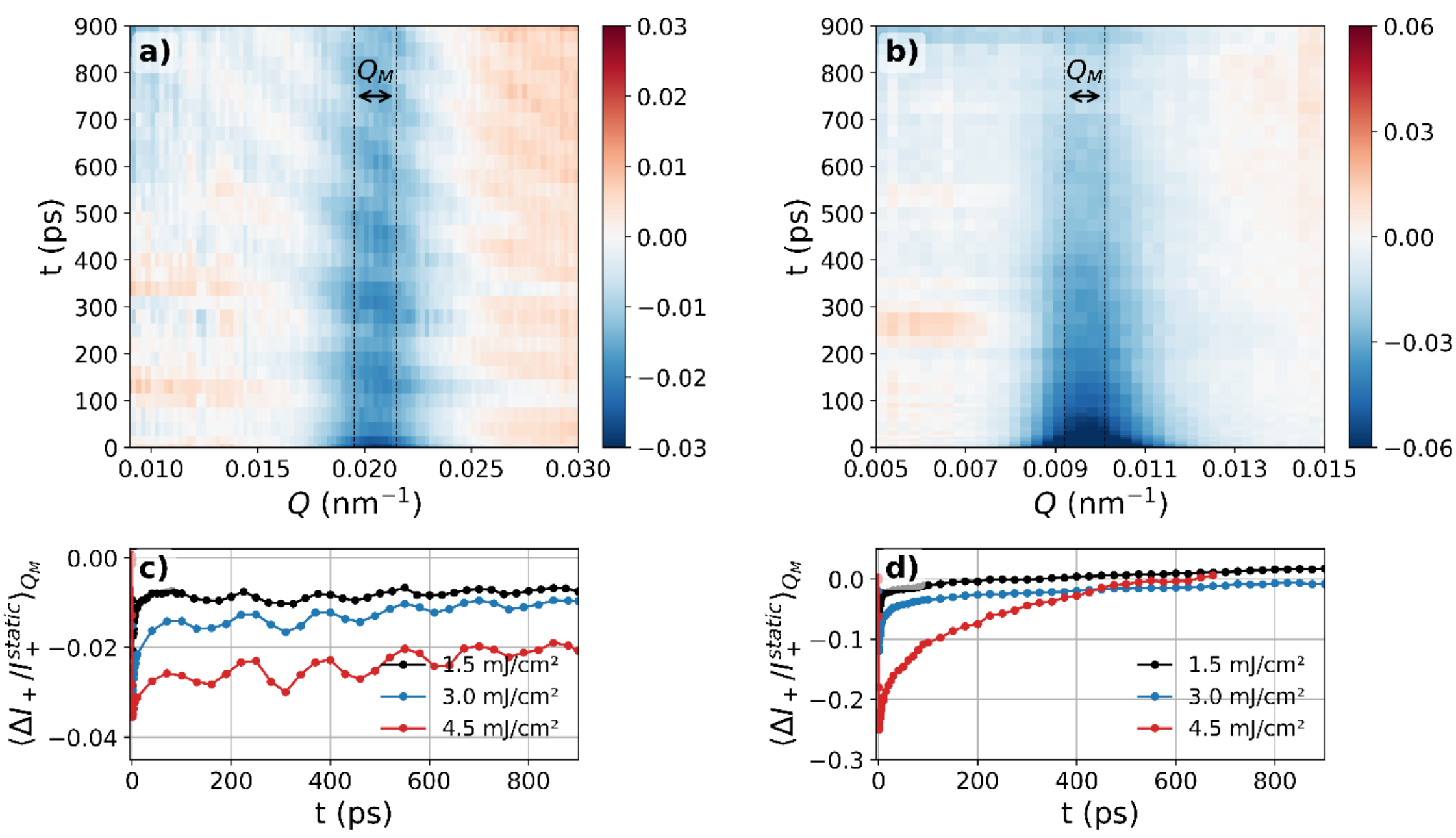

FIG. 2: **(a,b)** Time–momentum maps of $\Delta I_+/I_+^{static}$ at $F$ = 3.0 mJ cm$^{-2}$ for samples III (a) and I (b), respectively. **(c,d)** Corresponding momentum–integrated responses averaged around $Q_M$, shown as a function of pump–probe delay for $F$ = 1.5 (black), 3.0 (blue), and 4.5 mJ cm$^{-2}$ (red).

We expand the structural and magnetic amplitudes to first order as $f_0(t) = f_0 + \delta f_0(t)$ and $f_m(t) = f_m + \delta f_m(t)$ [8,9]. For the helicity-summed intensity $I_+$, the charge-magnetic interference term is odd under helicity reversal and cancels. The relative change of $I_+$ is

$$\frac{I_+^{pumped}(t) - I_+^{static}}{I_+^{static}} \approx \frac{2\Re e[f_0^* \delta f_0(t)]}{|f_0|^2 + |f_m|^2} + \frac{2\Re e[f_m^* \delta f_m(t)]}{|f_0|^2 + |f_m|^2} \tag{4}$$

The first term on the right-hand side of Eq. (4) represents a structural contribution associated with lattice displacements and strain. This contribution is weak and shows a small oscillatory component with an amplitude below ~0.5% over the studied fluence range. The oscillation period is in the tens to hundreds of picoseconds and does not depend strongly on pump fluence (see Fig. 2(a)), consistent with optically launched coherent surface phonons reported previously for metal films and multilayers [20]. The second term describes the magnetic contribution at $Q_M$, which is dominated by ultrafast demagnetization followed by recovery of magnetization [7,17].

For sample III ($n$ = 7), $\Delta I_+/I_+^{static}$ decreases by only about 2%–4% within the first picosecond, i.e., by only a few times more than the amplitude of the weak oscillatory structural contribution. Consequently, the surface-phonon-related oscillation remains visible on top of the magnetic background (Figs 2(a) and 2(c)). For sample I (n = 15), by contrast, $\Delta I_+/I_+^{static}$ exhibits a sharp and prompt decrease of approximately 10%–25%, depending on the pump fluence. In this case, the much stronger magnetic response obscures the weak structural modulation, so no clear structural oscillation can be identified in Figs. 2(b) and 2(d). A direct comparison between the two samples is further complicated by the different Q ranges probed in the experiment, which imply different oscillation periods, as well as by possible differences in surface roughness that can modify the structural scattering amplitude.

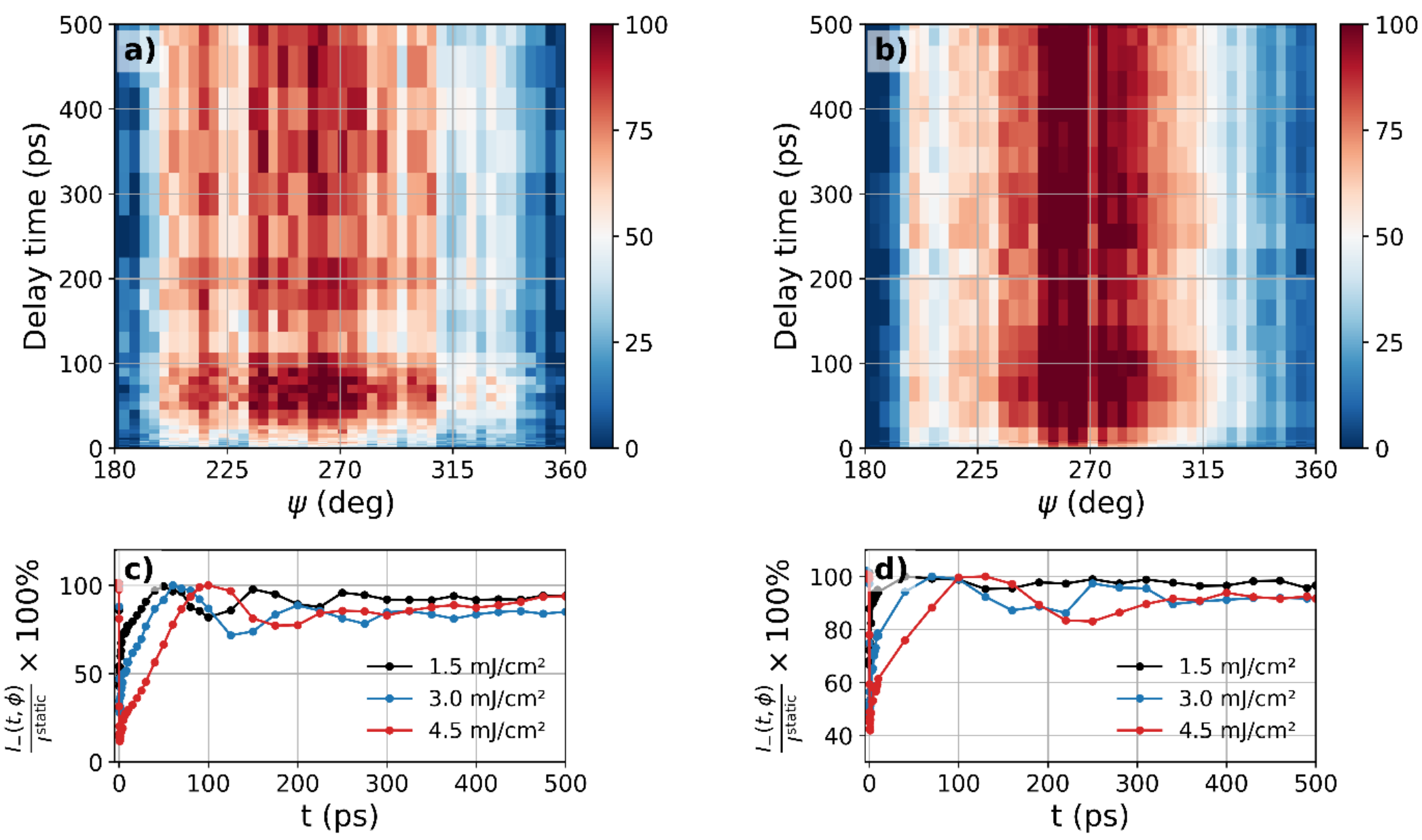

FIG. 3: **(a, c)** Time–azimuth maps of the normalized dichroic response $I_-(t,\psi)/I_-^{\text{static}}$, for samples I (a) and II (c) at a pump fluence of $F$ = 3.0 mJ cm$^{-2}$. **(b,d)** Corresponding azimuthally integrated transient dichroic responses shown as a function of pump–probe delay for pump fluences of 1.5 (black), 3.0 (blue), and 4.5 mJ cm$^{-2}$ (red).

Figs. 3(a) and 3(c) show time–azimuth maps of the normalized dichroic signal, $I_-(t,\psi)/I_-^{\text{static}}$, for samples I and II at a pump fluence of F = 3.0 mJ cm$^{-2}$, respectively. The maps were obtained by averaging over Q in the ranges [0.0092, 0.0101] nm$^{-1}$ around $Q_M$ for sample I and [0.0139, 0.0151] nm$^{-1}$ for sample II. The dichroic signal exhibits a pronounced maximum around ψ ≈ 270°, consistent with a Néel-type wall configuration, and this azimuth is used to track the chiral dynamics [9,10,12]. Note that the azimuthal range shown spans 180°–360°. Figs. 3(b) and 3(d) present the corresponding azimuthally integrated dichroic signals as a function of pump–probe delay for fluences F = 1.5, 3.0, and 4.5 mJ cm$^{-2}$. In both samples, the transient response displays damped oscillatory behavior over the first few hundred picoseconds. The oscillation amplitude decays with delay, so that only the first few maxima are well resolved. At the same time, the two samples show a clear difference in oscillation frequency.

Using the standard one-dimensional Walker domain-wall profile [13], the in-plane wall magnetization can be written as $m_{\parallel}(x,t) \propto M_s(t)\cos\varphi(t)\,\mathrm{sech}(x/\Delta(t))$, where $M_s$ is the saturation magnetization, $\Delta$ is the domain-wall width, and $\varphi$ is the internal wall angle. Its Fourier component at the stripe wave vector $Q_M$ yields, up to an overall numerical prefactor, a wall form factor proportional to $\Delta\,\mathrm{sech}(Q_M\Delta(t)/2)$. Since the dichroic XRMS signal is sensitive to the in-plane domain-wall magnetization [9], the dichroic scattering contribution can be expressed as

$$I_-(t) \propto M_s(t)\Delta(t)\cos\varphi(t)\,\mathrm{sech}\left(\frac{\pi Q_M\Delta(t)}{2}\right) \quad (5)$$

The dichroic signal is therefore sensitive to pump-induced changes in $M_s(t)$, $\Delta(t)$, and $\varphi(t)$ [9,12,21]. In particular, the cos $\varphi(t)$ factor makes $I_-(t)$ directly sensitive to transient deviations of the wall magnetization from the equilibrium Néel configuration.

The pump pulse drives the walls out of equilibrium and modifies their internal structure. We attribute the oscillatory component in $I_-(t)$ primarily to an internal domain-wall mode dominated by variations of $\varphi(t)$, with possible additional contributions from $\Delta(t)$ [12,13,21]. This interpretation is supported by the fact that the oscillation is pronounced in the dichroic signal but weak in the helicity-summed response. Periodic modulations of $M_s(t)$ and/or $\Delta(t)$ would, in general, also affect the non-dichroic signal, whereas the observed selectivity of the oscillation in $I_-(t)$ is more consistent with a dominant contribution from ϕ(t). This picture is also consistent with earlier observations of ultrafast transient distortions of chiral Néel walls toward mixed wall textures [12].

As a characteristic frequency scale for such a response, we use the scaling relation [13]

$$\omega_0(T) \sim \gamma\mu_0 H_k(T), \quad \mu_0 H_k(T) \propto \frac{K_{eff}(T)}{M_s(T)} \quad (6)$$

where *T* denotes the effective transient temperature reached after optical excitation, $\gamma$ is the gyromagnetic ratio, $H_k$ is the effective anisotropy field, and $K_{eff}$ is the effective perpendicular anisotropy. Equation (6) is used only as a scaling estimate for the characteristic frequency, since the exact eigenfrequency of an internal wall mode can also depend on i-DMI, the wall stiffness, and pinning [1,21]. With increasing pump fluence, transient heating modifies $K_{\mathrm{eff}}$ and $M_s$, thereby renormalizing the wall energy and the corresponding restoring torque [6,7,12]. The observed decrease in the oscillation frequency is therefore consistent with a reduction in the effective restoring field.

To quantify the chiral dynamics, the integrated dichroic traces were fitted over the full measured delay range. The non-oscillatory background was described using the three-exponential demagnetization and two-stage recovery response of Ref. [22], convolved with the experimental time resolution. The oscillatory modulation was represented phenomenologically by a cosine term with an exponentially decaying amplitude. The corresponding damping time was treated as an independent fit parameter and was not assigned a specific microscopic origin in the present analysis. Any possible time dependence of the oscillation frequency within the measured delay range (up to ~1 ns) was neglected. This procedure yielded a frequency f for each pump fluence *F*, and the resulting $f(F)$ values are summarized in Fig. 4 for samples I and II.

To parameterize the observed trend over the measured fluence range, the extracted $f(F)$ values were fitted empirically using a linear function, $f(F) = f_0 - kF$. We emphasize that this linear form was introduced only as a phenomenological description within the experimentally accessible fluence range, since the characteristic frequency is governed by fluence-induced changes in $K_{\mathrm{eff}}$ and $M_s$ and therefore is not expected to vary strictly linearly with *F*. The larger slope *k* obtained for sample I compared with sample II indicates a stronger fluence sensitivity of the wall-mode frequency in sample I. This difference likely reflects stack-dependent variations in the transient magnetic and thermal parameters that determine the restoring field. It may also be influenced by the larger equilibrium stripe-domain period in sample I, although the effect of domain size cannot be disentangled from the underlying magnetic parameters that set both the domain pattern and the wall dynamics.

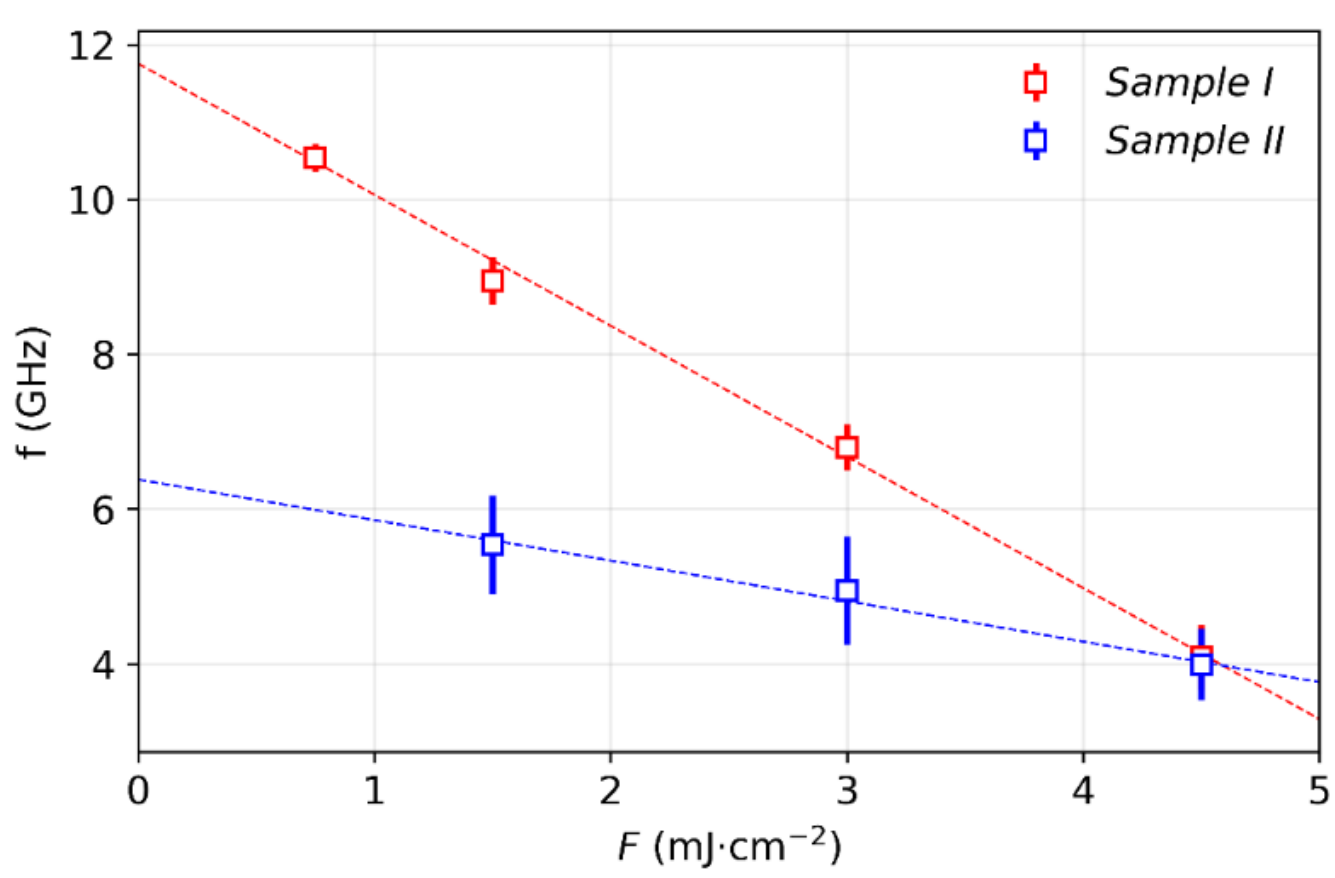

FIG.4: Fluence dependence of the extracted frequency for samples I (red) and II (blue). Markers show data and dashed lines show linear fits $f(F) = f_0 - kF$.

In summary, time-resolved XRMS resolves the ultrafast response of chiral domain walls in perpendicularly magnetized multilayers at the domain wavevector. The helicity-summed signal is dominated by demagnetization and recovery of the stripe-domain pattern, with only a weak additional oscillatory contribution consistent with optically generated surface phonons in the film stack. The dichroic signal, in contrast, shows a pronounced oscillatory response whose frequency decreases with pump fluence and therefore provides direct sensitivity to wall-specific dynamics. We interpret this softening as a transient reduction of the restoring forces that govern an internal domain-wall mode after optical excitation. The key result is that time-resolved dichroic XRMS separates wall-specific dynamics from the texture-averaged magnetic response and thereby provides reciprocal-space access to internal degrees of freedom of chiral domain walls in disordered labyrinth textures on picosecond time scales.

## ACKNOWLEDGMENTS

We thank the FERMI free-electron laser facility for beamtime allocated under Proposal No. 20199086.

### Funding

D.K. and C.G. acknowledge funding by the Deutsche Forschungsgemeinschaft (DFG) under Projects No. GU 535/9-1 and No. KS 62/3-1.

N.J. acknowledges financial support from the Agence Nationale de la Recherche (ANR), France, under grant agreements No.ANR-20-CE42-0012-01 (MEDYNA) and No.ANR-21-CE09-0042 (HYPNOSE), as well as from the France 2030 government investment plan managed by the French National Research Agency under grant No.ANR-22-EXSP-0008 (SPINCHARAC).

The team in Mainz acknowledges support from the DFG through the Collaborative Research Center TRR 173 (Spin+X; projects A01, B01, A07; project ID 268565370) and through the Elasto-Q-Mat (A12; project ID 422213477).

## AUTHOR DECLARATIONS

### Conflict of Interest

The authors have no conflicts to disclose.

## DATA AVAILABILITY

The data that support the findings of this study are available through the Virtual Unified Office (VUO) under proposal No. 20199086 (Ref. [23]).